\documentclass[aps,prl,twocolumn,groupedaddress,amsmath,epsfig,amssymb]{revtex4} % nofootinbib
\usepackage[utf8]{inputenc}
\usepackage[english]{babel}
\newtheorem{theorem}{Theorem}

\newtheorem{lemma}[theorem]{Lemma}
\usepackage{epsfig,amsmath,amssymb,bm,epsf,graphics,graphicx,psfrag,verbatim,subfigure,framed}
\usepackage{makecell}
\usepackage{bbm}
\usepackage{mathrsfs}
\usepackage{subfigure}
\usepackage{amsfonts}
\usepackage{color}
\usepackage{graphicx}% Include figure files
\newcommand{\xing}[1] {{\color{black} #1}}

\newcommand{\Imm}{{\boldsymbol I}}

\newcommand{\be}{\begin{equation}}
\newcommand{\ee}{\end{equation}}
\newcommand{\ba}{\begin{eqnarray}}
\newcommand{\ea}{\end{eqnarray}}

\begin{document}
	%\preprint{APS/123-QED}
	%Emergent structures
\title{Evading Thermodynamic Uncertainty Relations 
via Asymmetric Dynamic Protocols}
\author{Mingnan Ding$^{1}$}
\author{Chen Huang$^{1}$}
\author{Xiangjun Xing$^{1,2,3}$}
\email{xxing@sjtu.edu.cn}
\address{$^1$ Wilczek Quantum Center, School of Physics and Astronomy, Shanghai Jiao Tong University, Shanghai, 200240 China\\
$^2$ T.D. Lee Institute, Shanghai Jiao Tong University, Shanghai, 200240 China \\
$^3$ Shanghai Research Center for Quantum Sciences, Shanghai 201315 China}
	
	%\affiliation{}
	%\date{\today} %freeze this upon submission
	% but any date may be explicitly specifie
	%\pacs{82.70.Dd, 83.80.Hj, 82.45.Gj, 52.25.Kn}
	
	% [CHECK: need to add PACS numbers]
	%\pacs{61.41.+e, 61.43.Er, 62.20.Dc, 64.60.Ak, 64.60.Fr, 64.70.Dv}% PACS, the 
	%61.41.+e 61.43.-j
	%Physics and Astronomy
	
\begin{abstract}  
Many versions of Thermodynamic Uncertainty Relations  (TUR) have recently been discovered, which impose lower bounds on relative  fluctuations of integrated currents in irreversible dissipative processes, and suggest that there may be fundamental limitations on the precision of small scale machines and heat engines.  In this work we rigorously demonstrate that TUR can be evaded by using dynamic protocols that are asymmetric under time-reversal.  We illustrate our results using a model heat engine using two-level systems, and also discuss heuristically the fundamental connections between TUR and time-reversal symmetry. 
%\xing{FT: detailed FT or Crooks FT?}

% Of essential importance to the evading of TUR is the concept of {\em dominant path}, which takes almost all probability, and leads to vanishingly small fluctuations.  We show that symmetric dynamic protocols prohibit the existence of dominant path, and leads to lower bound of current fluctuations in small scale machines.  %The prohibition however can be evaded by using asymmetric protocols.  
 % 

\end{abstract}

\maketitle 
%:
%Significant progresses have been achieved in recent decades on non-equilibrium statistical physics at microscopic scales.  Among the new results, probably the most influential are  various versions of Fluctuation Theorems and work relations~\cite{Bustamante2005Nonequilibrium,Jarzynski-review,Seifert-review,Evans-Searles}, which not only reveal universal connections between time-reversal symmetry, fluctuations, and entropy productions in non-equilibrium processes, but also help shaping our understanding of small scale machines such as molecular motors, quantum heat engines, and thermo-electric junctions, which are often dominated by strong irreversibility and violent fluctuations~\cite{Kay2010,Rayment1993,Balzani2000,Douglas2012,Requicha2003,Pietzonka2016,Wang1998,Kolomelsky2007}.  

In recent years, several classes of inequalities have been discovered, which indicate trade-offs between various aspects of non-equilibrium processes.  For example, many versions of thermodynamic uncertainty relations (TUR)~\cite{Barato2015thermodynamic,Horowitz-2019-NP-review,Gingrich-2016,Pietzonka2016,Gingrich-2017,Koyuk-2019,Macieszczak-2018,Hasegawa-2019,Potts2019thermodynamic,Proesmans2017,Gingrich-2016-prl,proesmans2019hysteretic,Shiraishi-2017,Horowitz2017,Timpanaro-2019,Seifert-2017,vo2020unified,Pietzonka2018universal,Pal2019experimental,Friedman2020,Hasegawa2019quantum,Dechant2019}\cite{footnote-TUR} impose lower bounds of integrated current fluctuations in terms of entropy production, which signify a trade-off between dissipation and precision for small-scale machines.  Another example is the so-called speed-limit inequalities~\cite{Aurell-2011,Aurell-2012,Shiraishi-2016,Okuyama-2018,Shanahan-2018,Shiraishi-2018,Dechant-2019,Ito-2020}, which indicate a trade-off between efficiency and power output for general Markov processes.   Most recently, Dechant and Sasa~\cite{Dechant-2020} derived an upper bound for non-equilibrium response function in terms of fluctuations and relative entropy between the perturbed and reference states, which generated immediate interests~\cite{Dechant-2020-1,Hasegawa2019quantum,Dechant-2020-3}.  Given the large number of works published in recent years, it is highly desirable and urgent to understand whether these inequalities are consequences of more fundamental aspects of non-equilibrium physics, such as Fluctuation Theorems and Markovian property, or rather depend on specific system  details. 

The current situation of TUR is particularly vibrant and complex~\cite{Horowitz-2019-NP-review}.  The first version of TUR, which was proposed~\cite{Barato2015thermodynamic} and proved~\cite{Gingrich-2016} for continuous time Markov jump processes with local detailed balance, has the form: ${\langle \delta Q^2 \rangle}/{\langle Q \rangle^2}  \geq {2}/{\langle \Sigma \rangle }$,
where $Q$ is certain integrated current, and $\langle \Sigma \rangle $ the average entropy production.  It was very quickly generalized to other types of irreversible processes, including finite-time processes~\cite{Seifert-2017,Gingrich-2016-prl,Dechant2019,Horowitz2017}
discrete time processes~\cite{Shiraishi-2017,Proesmans2017}, and most recently  quantum processes~\cite{Hasegawa2019quantum,Friedman2020}.   Its relation to the linear response theory has also been clarified~\cite{Macieszczak-2018}.  \xing{More recently, Timpanaro  {\it et. al.} ~\cite{Timpanaro-2019} and Hasegawa {\it et. al.}~\cite{Hasegawa-2019} proved tighter TUR bounds in the form of ${\langle \delta Q^2 \rangle}/{\langle Q \rangle^2}  \geq f({\langle \Sigma \rangle })$ for general processes with symmetric dynamic protocols.  For small $\langle \Sigma \rangle $, $f({\langle \Sigma \rangle }) \sim 2/ \langle \Sigma \rangle $, so that the original TUR is restored.  For  large $\langle \Sigma \rangle $,  $f({\langle \Sigma \rangle })$ vanishes exponentially~\footnote{ Note that in the limit of large entropy production $\Sigma \rightarrow \infty$, both the original TUR bound in Refs.~\cite{Barato2015thermodynamic, Gingrich-2016} and the generalized bounds in Refs.\cite{Timpanaro-2019,Hasegawa-2019} vanish.  This is completely expected, since large $\langle \Sigma \rangle $ means either large system size or long time.  Either way, the law of large number come to play, so relative fluctuation of entropy production always reduces to zero.  In another word, various TUR bounds discovered for symmetric protocols are effective only for small size systems in short time.}.    
These generalized TUR  has their origin in Fluctuation Theorem~\cite{Seifert-review,Searles-1999,Crooks-1999}, and hence are universal and independent of model details. }The deep connection between fluctuation of entropy production and Fluctuation Theorem was previously studied by Merhav and Kafri~\cite{Merhav-2010}.

%Numerous counter-examples where TUR do not hold have also been discovered. 

%Note, however, the universal bounds found in these works agree with the original TUR bound only in the low dissipation limit. 

%These results are {\it universal}, i.e., independent of system details and valid for arbitrary non-equilibrium processes with symmetric dynamic protocols.  
%Their significance can be hardly overestimated. 
%These results are important because their validity is independent model details and therefore imposes universal bound on the precision of all microscopic machines that operate under time-reversal symmetric protocols.  
 
While almost all previous works concern processes with dynamic protocols that are symmetric under time-reversal, employment of asymmetric dynamic protocol provides another dimension for the issue of TUR.  Results about  asymmetric dynamic protocols are scarce, but did indicate a very different scenario.  Barrato and Seifert~\cite{Barato2016}  showed that Brownian clock driven periodically can achieve arbitrary high precision with arbitrary low dissipation.  Chun {\it et. al.} ~\cite{Barato2019-magnetic} showed that TUR may not work for system coupled to a magnetic field, which explicitly breaks the time-reversal symmetry.  In light of these results, it is highly desirable to know whether in principle TUR-bound can always be evaded via clever design of asymmetric dynamic protocols.   This question, which is independent of model details, is of fundamental  importance for study of small scale machines~\cite{Kay2010,Rayment1993,Balzani2000,Douglas2012,Requicha2003,Wang1998,Kolomelsky2007}.  

%Potts {\it et. al.}\cite{Potts2019thermodynamic} and Proesmans {\it et. al.}  ~\cite{proesmans2019hysteretic} recently derived an inequality involving current fluctuations both in the forward and in the backward processes. 

%In particular, the bound obtained by Timpanaro {\it et. al.}~\cite{Timpanaro-2019} is tight and is achieved.  

In this letter, we shall rigorously establish the absence of TUR-bound for processes with asymmetric dynamic protocols.  We shall minimize the relative current fluctuations with the constraints of (1) fixed entropy production and (2) satisfaction of Detailed Fluctuation Theorem, which is valid for processes with asymmetric protocols, and show that the lower bound for the current fluctuations is strictly zero. Hence all TUR-bounds can be evaded via clever designs of asymmetric dynamic protocols.  To illustrate our results, we also discuss a model of heat engine whose output fluctuation can be tuned arbitrarily small.  Finally we present a heuristic discussion in terms of Detailed Fluctuation Theorem why TUR-bounds exist in processes with symmetric dynamic protocols but not in those with asymmetric dynamic protocols.  \xing{Our results provide important insights for design of high precision microscopic machines and heat engines.  }

{\bf Fluctuation Theorem} \quad  Consider an irreversible stochastical process with path probability distribution $P_U[\gamma]$, \xing{where $U = \{\lambda(t), t \in (0,T)\}$ denotes the dynamic protocol, and $\lambda(t)$ is the time-dependent external parameter, and $(0,T)$ is the time-range of the process}.   Associated with this process is the backward process, with a reversed \xing{protocol $\bar U = \{\bar \lambda(t) = \lambda^*(-t), t \in (-T, 0)\}$ and a path probability distribution $P_{\bar U}[ \gamma]$. Note that $\lambda^*$ is related to $\lambda$ via reversal of odd parameters, such as magnetic field}.  Assuming that there is  separation of time-scale, and that all slow variables included in the model,  the entropy production $ \Sigma[\gamma]$ of the forward process along the path $\gamma$ satisfies~\cite{Maes2003time,seifert2005entropy}
\ba
{P_U[\gamma]} = {P_{\bar U}[\bar \gamma]} \, e^{ \Sigma[\gamma]},
\label{EPF-1}
\ea
which automatically implies $\Sigma[\gamma] =  - \bar \Sigma[ \bar \gamma]$, i.e., entropy production changes sign under the reversal of both the path and the dynamic protocol. \xing{Here $\bar \gamma$ is the time-reversal of the path forward $\gamma$, which is obtained from $\gamma$ via reversal of both time and odd variables.  Equation (\ref{EPF-1}) has been established for classical systems on very general ground~\cite{Maes2003time,seifert2005entropy}. It is also known to hold for certain quantum systems~\cite{Campisi-RMP-2011-QFT}}.  We shall be interested in  asymmetric protocols i.e., $U \neq \bar U$, which means that the forward and backward processes are macroscopically different.  
% For a system in contact with one or multiple heat baths, the total entropy production $\Sigma[\gamma]$ can be decomposed into two parts: a part due to entropy change of the baths, and a part due to the stochastic entropy of the system.  
%There are two typical scenarios where Eq.~(\ref{EPF-1}) becomes applicable:  (i) The system starts from an equilibrium state;  and (ii) the system is in a non-equilibrium steady state. 
 
Now consider a certain integrated current $Q[\gamma]$ as a functional of path, which is odd under time-reversal, i.e., $\bar Q[\bar \gamma] = - Q[\gamma]$.  It may be the amount of matter transported or  heat conducted along the path $\gamma$.  The joint probability density functions (pdf) of $\Sigma[\gamma]$ and $Q[\gamma]$ for the forward and backward processes are
\ba
p (\sigma, q)  &=& \sum_\gamma p_U[\gamma]\, 
\delta(\sigma - \Sigma[\gamma]) \delta (q - Q[\gamma]), \\
\bar p (\sigma, q)  &=&  \sum_\gamma p_{\bar U}[\gamma]\, 
\delta(\sigma - \bar \Sigma[\gamma]) \delta (q - \bar Q[\gamma]).  
\ea
Generalizing a theorem due to van der Broeck and Cleuren~\cite{Broeck-Cleuren-comment-2007}, we can readily obtain a generalized version of Detailed Fluctuation Theorem (DFT):
\ba
p (\sigma, q) = \bar{p}(-\sigma, -q) \, e^{\sigma}.  
\label{FT-1}
\ea
 All statistical properties of  $\Sigma$ and $Q$ for the forward and backward processes can be obtained from $p (\sigma, q) $ and $\bar{p} (\sigma, q)$.  
 If the dynamic protocol is symmetric, $\bar U = U$ and $\bar p = p$, and Eq.~(\ref{FT-1}) reduces to:
\be
p (\sigma, q) = {p}(-\sigma, -q) \, e^{\sigma},
\label{FT-sym}
\ee 
which was the starting point of studies in Ref.~\cite{Timpanaro-2019,Hasegawa-2019}.   Our aim is to minimize the variances $ \langle \delta Q^2 \rangle$ and $ \langle \delta \Sigma^2 \rangle$ under the constraint of fixed averages  $\langle \Sigma \rangle$ and $\langle Q \rangle$, for all distributions that satisfying Eq.~(\ref{FT-1}).   %Such a task was carried out in Ref.~\cite{Timpanaro-2019} for processes with symmetric protocol, for which Eq.~(\ref{FT-1}) reduces to Eq.~(\ref{FT-sym}).  

%\vspace{3mm}

{\bf Minimization of  Fluctuations} \quad   \xing{In this section we will show that DFT (\ref{FT-1}) implies no TUR bound for fluctuations.} Any continuous probability distribution can be approximated, to an arbitrary precision, by a discrete distribution.  Hence we only need to study discrete distributions.  We shall use the term {\em N-point distribution} to denote a probability distribution in the form of $p_N (\sigma, q) = \sum_i^N p_i \delta(\sigma - \sigma_i) \delta (q - q_i)$.  Let $\{p_N (\sigma, q), \bar{p}_N(-\sigma, -q) \}$ be a pair of $N$-point distributions for the forward and backward processes satisfying Eq.~(\ref{FT-1}).    As shown in detail in Supplementary Informations (SI), for any $N \geq 3$, we can always construct a pair of $(N-1)$-point distributions $\{ p_{N-1} (\sigma, q), \bar{p}_{N-1}(-\sigma, -q) \}$  which also satisfy Eq.~(\ref{FT-1}) but  has equal averages  $\langle \Sigma \rangle$ and $\langle Q \rangle$ and smaller fluctuations $ \langle \delta Q^2 \rangle$ and $ \langle \delta \Sigma^2 \rangle$.  Repeating this process, we eventually arrive at the following lemma: 
\begin{lemma} \quad
For any pair of N-point distributions $p_N(\sigma, q), \bar p_N(\sigma, q)$ obeying DFT Eq.~(\ref{FT-1}), we can always find a pair of 2-point distributions $p_2(\sigma, q), \bar p_2(\sigma, q)$ such that, comparing with $p_N(\sigma, q)$, $p_2(\sigma, q)$ has same averages $\langle \Sigma\rangle ,\langle Q\rangle$ and smaller variances $\langle \delta \Sigma^2 \rangle$ and $\langle \delta Q^2 \rangle$.   
\end{lemma} 
The detailed proof of this lemma is shown in SI. 
%The method of our proof is a substantial generalization of that used by Timpanaro et. al. \cite{Timpanaro-2019}, which was designed for processes with symmetric dynamic protocols.  

\xing{Let us further try to find the pair $p_2(\sigma, q), \bar p_2(\sigma, q)$ that produces the minimal fluctuations $\langle \delta \Sigma^2 \rangle$ and $\langle \delta Q^2 \rangle$.}  We shall first treat the simple case $Q = \Sigma$, so that  we only need to minimize $\langle \delta \Sigma^2 \rangle$ with fixed $\langle \Sigma \rangle$.   Let us write:
\begin{subequations}
\label{2-pts-pdf-def}
\ba
p_2(\sigma) &=& p \,  \delta(\sigma - \sigma_1) 
+   (1-p)  \, \delta(\sigma - \sigma_2) ,\\
\bar p_2 (\sigma) &=& p \, e^{-\sigma_1}\,  \delta(\sigma + \sigma_1) 
+   (1-p) \, e^{-\sigma_2} \, \delta(\sigma - \sigma_2), 
\quad \,\,
\ea
which satisfy Eq.~(\ref{FT-1}).   Whereas $p(\sigma)$ is already normalized, normalization of $\bar p(\sigma) $ fixes  $p$ in terms of $\sigma_1, \sigma_2$:
\ba
p =\frac{ 1- e^{- \sigma_2} }{e^{- \sigma_1} - e^{-\sigma_2}} .
%\quad 1-p= \frac{e^{- \sigma_1} -1}{e^{- \sigma_1} - e^{-\sigma_2}} . 
\label{p_1-p_2-solution}
\ea
\end{subequations}
 Using Eqs.~(\ref{2-pts-pdf-def}) we find averages and fluctuations of $\Sigma$ for the forward and backward processes:
\begin{subequations}
\label{var-sigma-2pts} 
\ba
\langle \Sigma \rangle &=& \frac{ \sigma_2 (e^{-\sigma_1} -1) 
+ \sigma_1 (1 - e^{-\sigma_2})}
{e^{- \sigma_1} - e^{- \sigma_2} }, \\
\langle \delta \Sigma^2 \rangle &=& 
\frac{(\sigma_1 - \sigma_2)^2 (e^{-\sigma_1} -1) (1- e^{- \sigma_2})}
{( e^{- \sigma_1} - e^{- \sigma_2})^2 }, \\
\overline{\langle \Sigma \rangle} &=& \frac{ - \sigma_1 e^{-\sigma_1} (1 - e^{-\sigma_2})
- \sigma_2  e^{- \sigma_2} (e^{-\sigma_1} -1) }
{e^{- \sigma_1} - e^{- \sigma_2} },  \quad \\
\overline{ \langle \delta \Sigma^2 \rangle} &=& 
\frac{(\sigma_1 - \sigma_2)^2 (e^{\sigma_2} -1) (1- e^{\sigma_1})}
{( e^{\sigma_1} - e^{ \sigma_2})^2 }.
\ea
\end{subequations}

If $\sigma_1 = 0$ ($\sigma_2 = 0$), $ p = 1$ ($p=0$), the entropy production is always zero, which corresponds to a reversible process.  This is not what we aim to study in this work, hence we shall assume that neither of $\sigma_1, \sigma_2$ vanish.  From Eq.~(\ref{p_1-p_2-solution}) we see to make $p$ or $1-p$ both positive, $\sigma_1, \sigma_2$ must have different signs.  Let us assume $\sigma_1<0 < \sigma_2$. 

 Let us now fix $\sigma_2 >0$, and make $| \sigma_1| = - \sigma_1 \gg 1$, hence $e^{- \sigma_1} \gg 1$, and   $p \sim e^{\sigma_1} = e^{- |\sigma_1|} \ll 1$, according to Eq.~(\ref{p_1-p_2-solution}). From Eqs.~(\ref{var-sigma-2pts}) we obtain the following asymptotics:
\begin{subequations}
\label{Sigma-asymp}
\ba
\langle \Sigma \rangle  &\sim& \sigma_2,
\label{Sigma-asymp-1}
\\
\langle \delta \Sigma^2 \rangle &\sim& 
\sigma_1^2 (1- e^{- \sigma_2}) e^{\sigma_1}
\rightarrow 0,
\label{Sigma-asymp-2}
\\
{\langle \delta \Sigma^2 \rangle} /{\langle \Sigma \rangle^2}
&\sim& \sigma_2^{-2}\sigma_1^2  
(1- e^{- \sigma_2}) e^{\sigma_1} 
\rightarrow 0. \quad\quad
\label{Sigma-asymp-3}
\ea
Hence in the forward process, $\langle \Sigma \rangle$ converges to a finite limit, whereas $\langle \delta \Sigma^2 \rangle $ becomes exponentially small in $\sigma_1$.  Similarly for the backward process we obtain
\ba
\overline{\langle \Sigma \rangle} &\sim& - \sigma_1 
+ \sigma_1 e^{- \sigma_2}
\rightarrow \infty, 
\label{Sigma-asymp-4}\\
% \rightarrow + \infty, \\
\overline{ \langle \delta \Sigma^2 \rangle} &\sim& \sigma_1^2
 e^{- 2 \sigma_2} (e^{\sigma_2} -1)
 \rightarrow \infty, 
 \label{Sigma-asymp-5}
\\
% \rightarrow + \infty.
\overline{\langle \delta \Sigma^2 \rangle} /
 \overline{\langle \Sigma \rangle}^2
 &\sim& {1}/{(e^{\sigma_2} -1)}.
 \label{Sigma-asymp-6}
\ea
\end{subequations}
Hence in the backward process, both $\langle \Sigma \rangle$ and $\langle \delta \Sigma^2 \rangle $  diverge, whereas $\overline{\langle \delta \Sigma^2 \rangle} / \overline{\langle \Sigma \rangle}^2$ converges to a finite limit.  

Let us now consider the general case where $Q$ is an integrated current distinct from entropy production.  The joint distributions of  $\Sigma$ and $Q$ have the following forms:
\begin{subequations}
\label{2-pts-pdf-def-joint}
\ba
p_2(\sigma, q) &=& p \,  \delta(\sigma - \sigma_1) \delta(q - q_1) 
\nonumber\\
&+&   (1-p)  \, \delta(\sigma - \sigma_2)  \delta(q - q_2) ,
\\
\bar p_2 (\sigma, q) &=& p \, e^{-\sigma_1}\,  \delta(\sigma + \sigma_1) 
 \delta(q + q_1) 
\nonumber\\
&+&   (1-p) \, e^{-\sigma_2} \, \delta(\sigma + \sigma_2)
 \delta(q + q_2) , 
\quad
\ea
\end{subequations}
where $p$ is still given by Eq.~(\ref{p_1-p_2-solution}).  Whilst the average entropy productions are still given by Eqs.~(\ref{var-sigma-2pts}), the average and variance of $Q$ can be calculated using Eqs.~(\ref{2-pts-pdf-def-joint}).  Recall that we fix $\sigma_2 >0$ and let $\sigma_1 \rightarrow - \infty$.  In this limit, we have for the forward process: 
\begin{subequations}
\label{Q-Q^2-for-back}
\ba
\langle \delta Q^2 \rangle &\sim &
{(q_1 - q_2)^2 e^{\sigma_1} (1- e^{- \sigma_2})}, \\
\langle Q \rangle & \sim& q_2. 
\ea
For all reasonable physical models, we expect that entropy production scales quadratically (or at least polynomially) with the current.  Hence $q_1$ diverges with $\sigma_1$, whereas $q_2$ remains finite.  From Eqs.~(\ref{Q-Q^2-for-back}) we find that $\langle \delta Q^2 \rangle$ converges to zero, whereas $\langle Q \rangle $ remains fixed.  The relative fluctuation of $Q$ converges to zero:
\be
\langle \delta Q^2 \rangle / \langle Q \rangle^2 
\sim  q_2^{-2} q_1^2  
(1- e^{- \sigma_2}) e^{\sigma_1} 
\rightarrow 0.  
\ee 
For the backward process we have 
\ba
\overline{ \langle \delta Q^2 \rangle} &\sim &
{(q_1 - q_2)^2 e^{ - \sigma_2} (1- e^{- \sigma_2})},  \\
\overline{ \langle Q \rangle }
 & \sim& - (1 - e^{- \sigma_2}) q_1 - e^{- \sigma_2} q_2. 
\ea
hence both $\overline{ \langle \delta Q^2 \rangle} $ and  $\overline{ \langle Q \rangle }$ diverge with $q_1$.  The relative fluctuation of $Q$ converges to the same limit as in Eq.~(\ref{Sigma-asymp-6}):
\be
\overline{ \langle \delta Q^2 \rangle} /\overline{ \langle Q \rangle } ^2 
\sim  {1}/{(e^{\sigma_2} -1)}, 
\ee
\end{subequations}

Naturally one may wonder whether it is possible to make current fluctuations small for both the forward and backward process.  This is impossible.  In Refs.~\cite{Potts2019thermodynamic,proesmans2019hysteretic} the following inequality was derived from DFT (\ref{FT-1}):
\ba
\frac{\langle \delta Q^2 \rangle + \overline{\langle \delta Q^2 \rangle} }
{( \langle Q \rangle + \overline{ \langle Q \rangle})^2}
 \geq { {\rm exp}-  \frac{1}{2}\left[ \langle \Sigma \rangle+
 \overline{\langle \Sigma \rangle} \right]}.  
 \label{TURasy}
\ea
If $\langle Q \rangle, \overline{\langle Q \rangle} $, $\langle \Sigma \rangle, \overline{\langle \Sigma \rangle} $ are all finite, and $\langle \delta Q^2 \rangle, \overline{\langle \delta Q^2 \rangle} $ are very small, Eq.~(\ref{TURasy}) must be violated.  On the other hand, for the two-point distributions we studied above, using Eqs.~(\ref{Q-Q^2-for-back}), we see that the LHS of  Eq.~(\ref{TURasy}) converges to $ {1}/{(e^{\sigma_2} -1)} > 0$, whereas RHS of Eq.~(\ref{TURasy}) converges to zero.  Hence the inequality (\ref{TURasy}) is not tight.  

\xing{These results show unambiguously that there is no TUR bound for entropy production as long as asymmetric dynamic protocols are allowed.  In Ref.~\cite{Brandner-2018}, Brandner et. al. studied an irreversible process with magnetic field and certain dynamic protocol, and discovered a non-vanishing TUR bound for current fluctuations.  This does not contradict our theory.  Instead our theory means that the TUR bound discovered in Ref.~\cite{Brandner-2018} can be evaded if more sophisticated protocols are used. }

%The lesson we learn here is that we can use asymmetric dynamic protocol to evade TUR in the forward or the backward process, but not in both processes.  

\begin{figure}[t!]
	\centering
	\includegraphics[height=1.8in]{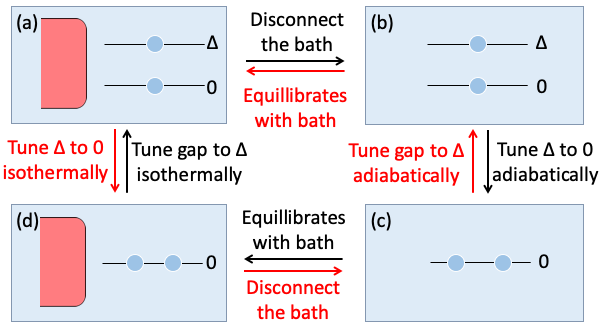}
	\caption{Schematics of the qubit model.  The red regions denote heat baths. Black arrows and red arrows denote respectively forward and backward dynamic protocols. 
	Two energy levels are degenerate in (c) and (d).}
	\label{fig:process}
	\vspace{-3mm}
\end{figure}

\vspace{2mm}
{\bf A model heat engine} \quad  \xing{ We shall now discuss a concrete irreversible process with finite $\langle \Sigma \rangle$ and vanishingly small current fluctuations. } Consider a two-level system with a Hamiltonian 
\be
 \hat H (\Delta) = \Delta  \,  \begin{pmatrix} 1 & 0\\ 0 & 0 \end{pmatrix},
 \quad  \Delta >0,
%\quad \hat \sigma = \begin{pmatrix} 1 & 0\\ 0 & -1 \end{pmatrix}. 
\ee 
where the energy of the excited state $\Delta$ that can be tuned externally.  We shall construct an irreversible cycle that starts from a Gibbs state $\hat \rho (\Delta)  = Z^{-1} e^{- \beta \hat H(\Delta) }$, illustrated by panel (a) of Fig.~\ref{fig:process} (top left), which can be obtained by connecting the system to a heat bath with temperature $T = 1/\beta$.  As illustrated by the black arrows in Fig.~\ref{fig:process}, the cycle consists of the following four steps $ (a) \stackrel {\tiny 1} {\rightarrow}  (b) \stackrel {\tiny 2 } {\rightarrow} 
 (c) \stackrel {\tiny 3 } {\rightarrow}  (d) \stackrel {\tiny 4} {\rightarrow}  (a)$:
\begin{enumerate}
\item {\em Disconnection} of the system from the heat bath.  
\item {\em Adiabatic process}:  We adiabatically tune the parameter $\Delta$ to $0$, during which the system remains in the state $\hat \rho = Z^{-1} e^{- \beta \hat H(\Delta) }$.  
\item {\em Equilibration process}:  We reconnect the system to the bath, and the system re-equilibrates to a new Gibbs state $\hat \rho = Z^{-1} e^{- \beta \hat H(0) } = \hat \Imm/2$.  This step is irreversible process with an entropy increase.
\item {\em Isothermal process}: With the system connected to the bath, the gap is tuned isotatically back to $\Delta$ so that the density matrix returns to the initial state $\hat \rho (\Delta) = Z^{-1} e^{- \beta \hat H(\Delta) }$.  
  %(The work done is actually a stochastic variable.  In the isostatic limit, however, it converges to a deterministic value.  )
\end{enumerate}
\xing{In stages (a) and (d), the system is in the relevant Gibbs-Boltzmann state.  In stages (b) and (c), the system is either at the excited level (with probability $p = e^{- \beta \Delta}/(1+ e^{- \beta \Delta})$) or at the ground level (with probability $1-p = 1/(1+ e^{- \beta \Delta})$). There are two paths for the adiabatic process, $\gamma_1  =  \{{\rm Gibbs} \rightarrow {\rm excited} \rightarrow  {\rm excited}  \rightarrow {\rm Gibbs}\}$ with probabilities $p$,  and $\gamma_2 =  \{{\rm Gibbs} \rightarrow {\rm ground} \rightarrow  {\rm ground}  \rightarrow {\rm Gibbs}\} $, with probabilities $1-p$.  These two paths merge in stages (a) and (d). }

The only integrated current in this problem is the entropy production.  Since the process is cyclic,  the entropy production is related to the work done by the external force via $\Sigma = \beta W$.  Let us consider the adiabatic process.  Along the path $\gamma_1$, the system remains in the excited state, and the external force does work $ - \Delta$.  Along the path $\gamma_2$, the work is identically zero.  During the isothermal process, two path coincide, and the work done equals to the change of system free energy, which is $T \log {2}/( {1 + e^{- \beta \Delta}})$.   Hence the total works along two paths are respectively:
\begin{subequations}
\ba
W[\gamma_1] &=& - \Delta + T \log \frac{2}{1 + e^{- \beta \Delta}}
= T \log (2 p), \\
W[\gamma_2]  &=&  T \log \frac{2}{1 + e^{- \beta \Delta}} = T \log 2 (1-p). 
\ea
\end{subequations}
Hence the path probabilities and the entropy production of $\gamma_1, \gamma_2$ are respectively
\begin{subequations}
\label{example-details}
\ba
&& p_1 = p =\frac{ e^{- \beta \Delta}}{ 1+ e^{- \beta \Delta} }, 
\quad p_2 = 1-p,\\
&& \sigma_1 = \log (2 p), \quad 
\sigma_2 = \log 2(1-p).
\ea
This is a special case of Eqs.~(\ref{2-pts-pdf-def}), with only one independent parameter $p$ (or equivalently $\beta \Delta$). 

%which satisfy Eq.~(\ref{p_1-p_2-solution}).  As we tune the energy gasp to be large, such that $ p \rightarrow 0$, and the 
% Comparing with Eqs.~(\ref{2-pts-pdf-def}), we see that

The backward process is obtained by reversing the dynamic protocols, shown as all red arrows in Fig.~\ref{fig:process}.  In the backward process, the system starts from the equilibrium state $\hat \rho = Z^{-1} e^{- \beta \hat H(\Delta) }$ (a), and goes through consecutively isothermal, disconnection, and adiabatic, and connection steps~\footnote{Note the reversed operations of disconnection/connection with the bath is connection/disconnection with the bath.  For example, in the forward process, the system gets disconnected from the bath with $\Delta >0$, whereas in the backward process, disconnection happens when $\Delta = 0$.  This shows explicitly that the protocol is indeed asymmetric under time-reversal.  }.  The two paths diverge from each other during the adiabatic process (c)  $\rightarrow$ (b).  In the stage (c), two levels are degenerate, and the system is in each of them with probability $1/2$.   The works done by the external force during each path in the backward process are the negative of those in the forward process.  Hence we have for the backward process:
\ba
&& \bar p_1 = \frac{1}{2}, \quad \bar p_2 = \frac{1}{2},\\
&&\bar  \sigma_1 = - \log (2 p), \quad 
\bar \sigma_2 = - \log 2(1-p).
 \label{p-bar-1-2}
\ea
\end{subequations}
We easily verify that Eqs.~(\ref{example-details}) satisfy the entropy production formula,  Eq.~(\ref{EPF-1}).  

As we make $\beta \Delta$ large, $p \rightarrow 0$, and $\sigma_1 \rightarrow 0$, $\sigma_2 \rightarrow \log 2$, and we recover all the asymptotics in Eqs.~(\ref{Sigma-asymp}).  More concretely, in the forward process, $\langle \Sigma \rangle \rightarrow \log 2$, and $\langle \delta \Sigma^2 \rangle \sim (\beta \Delta)^2 e^{ - \beta \Delta} \rightarrow 0$, whereas in the backward process, we have $ \overline{ \langle \Sigma \rangle} \rightarrow \beta \Delta / 2$, and $\overline{ \langle \delta \Sigma^2 \rangle }
\sim (\beta \Delta)^2/2$.  The relative fluctuation of entropy production of the backward process converges to a finite limit $\overline{ \langle \delta \Sigma^2 \rangle }/  \overline{ \langle \Sigma \rangle} ^2 \rightarrow 2$.  

%In this toy model, there are only two paths.  In the limit of large gap $\beta \Delta \rightarrow \infty$, and $p \rightarrow 0$, hence the path $\gamma_2$ takes almost all the probability, and $\gamma_1$ becomes practically improbable.  As a consequence, the fluctuation of output becomes negligible.  By contrast, in the backward process, two parts are equally probably regardless of the value of $\Delta$, and hence the fluctuations of output remain non-negligible.  Evidently, the fact that the distributions of path probability for the forward and the backward processes are different plays essential role here.  

\begin{figure}[t!]
	\centering
	\includegraphics[height=1.0in]{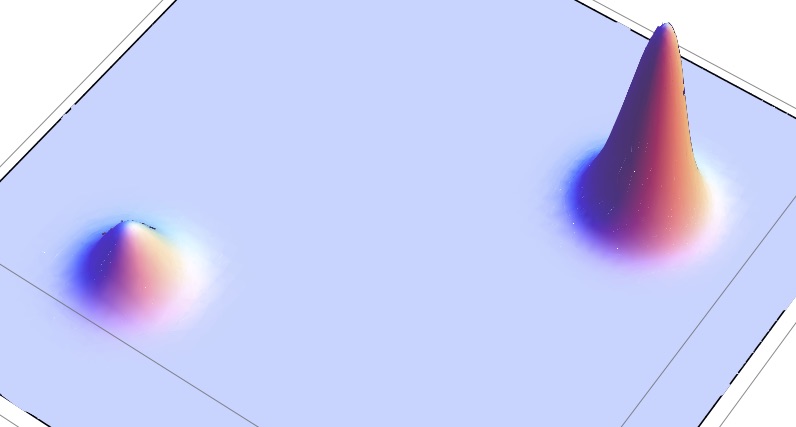}
	\caption{A cartoon of pdf $p(\sigma, q)$ for processes with symmetric dynamic protocol.  The high peak in the right is centered at $(q_0, \sigma_0)$, whereas the mirror peak in the left, which is dictated by the Detailed Fluctuation Theorem, is located at $(-q_0, -\sigma_0)$.  The mirror peak can be diminished by using asymmetric dynamic protocols, thereby evading TUR bounds. }
	\label{fig:peaks}
	\vspace{-5mm}
\end{figure}

%\vspace{2mm}
{\bf Heuristic discussion} \quad We conclude our work with a heuristic discussion about TUR bounds in general irreversible processes.  We aim, via tuning of dynamic protocols, to reduce the fluctuations of $\Sigma$ and $Q$, {with their averages fixed and finite.}  The distribution $p(\sigma, q)$ then should have significant values only near a single point $(\sigma_0,q_0)$.  Everywhere else $p(\sigma, q)$ must be negligibly small.  However,  if the dynamic protocol is symmetric under time-reversal, Eq.~(\ref{FT-sym}) must be respected, according to which there is a peak at the mirror point $(-\sigma_0, -q_0)$, as illustrated in Fig.~\ref{fig:peaks}.    Equation (\ref{FT-sym}) also dictates that the probabilities of the peak and of the mirror peak are respectively $1/(1+ e^{- \sigma_0})$ and $1/(1 + e^{\sigma_0})$.  Using these probabilities one easily  obtain  $\langle \Sigma \rangle = \sigma_0 \tanh (\sigma_0/2)$ and $\langle \delta Q^2 \rangle /\langle Q\rangle^2 = 2 \, {\rm csch}^2(\sigma_0 /2)$.  The latter is in fact exactly the sharp TUR bound obtained recently by Timpanaro  {\it et. al.}~\cite{Timpanaro-2019}.  For processes with asymmetric dynamic protocols, however, the relevant Detailed Fluctuation Theorem is Eq.~(\ref{FT-1}) rather than Eq.~(\ref{FT-sym}).   But Eq.~(\ref{FT-1}) does not imposes any constraint on $p(\sigma, q)$. Instead, it determines $\bar p(\sigma, q)$ in terms of $p(\sigma, q)$.  Hence by using asymmetric protocols,  the mirror peak can be arbitrarily diminished, and the TUR bound can   be evaded, with average entropy production finite.   Note that this heuristic discussion is applicable to models with continuous distributions of dynamic paths as well.

{\bf Acknowledgement}\quad 
 X.X. acknowledges support from NSFC via grant \#11674217, as well as additional support from a Shanghai Talent Program.  This research is also supported by Shanghai Municipal Science and Technology Major Project (Grant No.2019SHZDZX01).

% Information ratchet:

%\bibliography{ref_new.bib}

%\pagebreak

%The Brownian motion in liquid has also been studied \cite{kheifets2014observation, huang2011direct}. It is characterized not only by much faster timescales but also by extra force terms due to high density and viscosity of the liquid to correctly describe the  interplay between the particle and the medium. Thus Brownian motions in liquid are out of our consideration.

%The radius of the Brownian particle is typically $10^{-9}m<a<5\times10^{-7}m$. In a colloidal system, there are in general three different timescales: the short atomic scale $\tau_s\approx 10^{-12}s$, the relaxation time of the Brownian particle velocity $\tau=\frac{m}{\gamma}\approx10^{-3}s$, and the time the particle diffuses its own radius $\tau_r=a^2/D$. In dense colloidal suspensions $\tau_r$ can be minutes or even hours.

%Note the equation $\langle\Delta x^2\rangle=2Dt$ is only valid
%when $t\gg\tau$, i.e., in the diffusive regime. At very short time scales $t\ll\tau$, the motion is ballistic, and the MSD is predicted to be $\langle\Delta x^2\rangle=\frac{kT}{m}t$.

%\appendix

\pagebreak

\pagebreak

\begin{center}
{\bf Supplementary Informations}
\end{center}
\section*{Proof of Lemma 1}

Let us write down a pair of N-point probability distributions, which satisfy the Detailed Fluctuation Theorem (\ref{FT-1}):
\begin{subequations}
\ba
p_N (\sigma, q) &=& \sum_{i = 1}^N p_i \, \delta(\sigma - \sigma_i) \delta(q - q_i), \\
\bar p_N (\sigma, q) &=& \sum_{i = 1}^N p_i \,e^{ - \sigma_i} \, 
\delta(\sigma + \sigma_i) \delta(q + q_i).
\ea
\end{subequations}
Normalization of two probability distributions require:
\begin{subequations}
\label{normalization-pdf}
\ba
\sum_{i = 1}^N p_i &=& 1, \\
\sum_{i = 1}^N p_i \, e^{ -\sigma_i} & =& 1.
\ea
\end{subequations}
Let $\sigma_1$ and $\sigma_2$ be the smallest and biggest of $\{\sigma_i, i = 1, \cdots, N \}$, and let us choose another point $\sigma_3$ such that $\sigma_1 < \sigma_3 < \sigma_2$.  (If two or all of $\sigma_1, \sigma_2, \sigma_3$ are equal, we can tune $\sigma_1, \sigma_2, \sigma_3$ as well as $p_1, p_2, p_3$ such that $\sigma_1 < \sigma_3 < \sigma_2$ and at the same time the normalization conditions (\ref{normalization-pdf}) remain valid, and $\langle \Sigma \rangle$ and $\langle Q \rangle$ remain fixed.  This is always possible since we have six parameters to satisfy four independent constraints.  Furthermore by choosing the direction of variation, we can always make sure that the variances of $\Sigma$ and $Q$ do not increase.) Now let us write down $\langle \Sigma \rangle, \langle \Sigma^2 \rangle, \langle Q \rangle, \langle Q^2 \rangle$:
\ba
\langle Q \rangle &=& p_1 q_1 + p_2 q_2 + p_3 q_3 + \sum_{i >3} p_i q_i ,
 \label{average-Q} \\
\langle Q^2 \rangle &=& p_1 q_1^2 + p_2 q_2^2 + p_3 q_3^2
+ \sum_{i >3} p_i q_i^2,
 \label{average-Q^2} \\
\langle \Sigma \rangle &=& p_1 \sigma_1 + p_2 \sigma_2 
+ p_3 \sigma_3 + \sum_{i >3} p_i \sigma_i ,
 \label{average-sigma} \\
\langle \Sigma^2 \rangle &=& p_1 \sigma_1^2 + p_2 \sigma_2^2 + p_3 \sigma_3^2
+ \sum_{i >3} p_i \sigma_i^2. 
 \label{average-sigma^2}
\ea

\begin{figure*}[t!]
	\centering
	\includegraphics[height=1.8in]{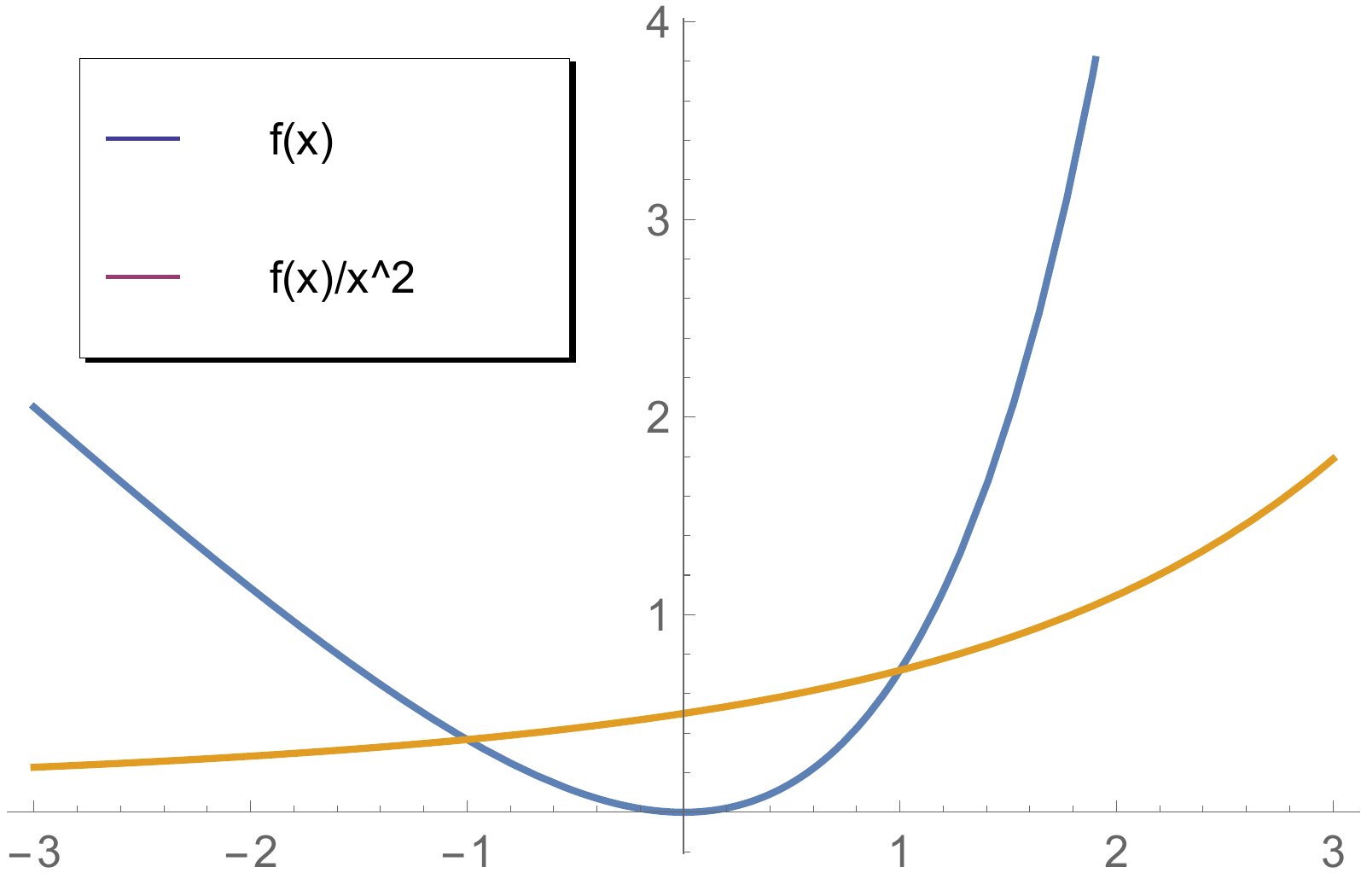}
	\caption{Plots of functions $f(x) = e^{x} -1- x$ and $f(x)/x^2$.}
	\label{fig:functions}
	%\vspace{-3mm}
\end{figure*}

We shall redistribute infinitesimally the probability $p_3$ to $p_1, p_2$, and at the same time tune $\sigma_3$ such that (a) Eqs.~(\ref{normalization-pdf}) remain valid, and (b) $\langle \Sigma \rangle$ remains invariant.   Let us consider the infinitesimal transformation:
\ba
&& p_i \rightarrow p_i + d p_i, \quad i = 1,2,3,\\
&& \sigma_3 \rightarrow \sigma_3 + d \sigma_3 ,
\ea
whereas all other $p_k,\sigma_k$ remain invariant.  Let us take the differential of Eqs.~(\ref{normalization-pdf}) and (\ref{average-sigma}):
\ba
dp_1 + dp_2 + dp_3 &=&0, \\
e^{- \sigma_1} dp_1 + e^{- \sigma_2} dp_2 +e^{- \sigma_3} dp_3 
- e^{- \sigma_3} p_3 d \sigma_3 &=&0, \quad \\
d \langle \Sigma \rangle = \sigma_1 dp_1 + \sigma_2 dp_2 + \sigma_3 dp_3 + p_3 d \sigma_3 &=&0.  \quad
\ea
Using these equations we can express $dp_2, dp_3, d\sigma_3$ in terms of $dp_1$.  In particular we have 
\ba
d p_2 = - \frac{f(\sigma_3 - \sigma_1)}{f(\sigma_3 - \sigma_2)} dp_1,
\ea
where  $f(x)$ is  a concave function with positive second derivative: 
\be
f(x) = e^{x} - 1 - x.  
\ee
As illustrated in Fig.~\ref{fig:functions}, $f(x)$ is non-negative and achieves minimum $f = 0$ at $x = 0$.  Hence $dp_2$ and $dp_1$ always have different signs, since $\sigma_1 < \sigma_3 < \sigma_2$. 

We can then calculate the differential of $\langle \Sigma^2 \rangle$ and express in terms of $dp_1$:
\ba
d \langle \Sigma^2 \rangle  
%&=& 
%\frac{ (\sigma_3 - \sigma_1)^2 f(\sigma_3 - \sigma_2)
%- (\sigma_3 - \sigma_2)^2 f(\sigma_3 - \sigma_1) }
%{f(\sigma_3 - \sigma_2) } dp_1 
%\nonumber\\
&=& \frac{ (\sigma_3 - \sigma_1)^2(\sigma_3 - \sigma_2)^2 }
{f(\sigma_3 - \sigma_2)}
\nonumber\\
&\times& \left[ \frac{f(\sigma_3 - \sigma_2)} {(\sigma_3 - \sigma_2)^2}
-  \frac{f(\sigma_3 - \sigma_1)} {(\sigma_3 - \sigma_1)^2} \right] dp_1 .
\label{difference-RHS}
\ea
As illustrated in Fig.~\ref{fig:functions}, the function $f(x)/x^2$ is strictly monotonically increasing.  Since $\sigma_3 - \sigma_2<0$ and $\sigma_3 - \sigma_1> 0$, the difference in the bracket in the RHS of Eq.~(\ref{difference-RHS}) is strictly negative.   We shall chose $dp_1 >0$, then we have $dp_2 <0$,  and $d \langle\Sigma^2 \rangle <0$, and hence the variance $ \langle \delta \Sigma^2 \rangle$ decreases.  

\begin{figure*}[ht!]
	\centering
	\includegraphics[height=2.2in]{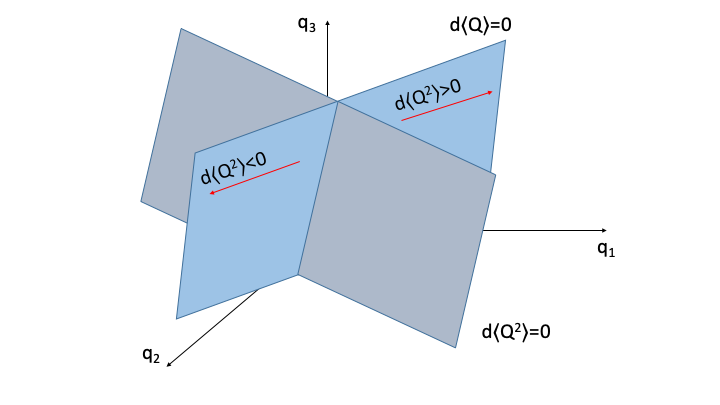}
	\caption{Illustration of the redistribution process to reduce the variance of Q.}
	\label{fig:aveq}
	%\vspace{-3mm}
\end{figure*}

Let us now take care of the other variable $Q$.  Let us take the differential of Eq.~(\ref{average-Q}) and set it to zero:
\ba
d\, \langle Q \rangle = \sum_{i = 1}^3 p_i d q_i + \sum_{i = 1}^3  q_i d p_i  = 0.
\label{d-Ave-Q=0}
\ea
With $dp_2, dp_3$ determined previously in terms of $d p_1$, this equation defines a plane in the three dimensional space spanned by $dq_1, dq_2, dq_3$.  Now let us take the differential of Eq.~(\ref{average-Q^2}) and set it to zero:
\be
d \langle Q^2 \rangle %= d \left(\sum_i p_i q_i^2 \right) 
= \sum_i 2 p_i q_i d q_i +  \sum_i  q_i^2 d p_i = 0, 
 %(q_1^2 - q_3^2 ) dp_1 + (q_2^2 - q_3^2) dp_2
%+ 2 p_1 q_1 d q_1 + 2 p_2 q_2 d q_2.
\label{dAve-Q^2}
\ee 
which defines another plane in the same space.  Assuming $q_1, q_2, q_3$ are not all identical, (If this condition is not satisfied, we can adjust $q_1, q_2, q_3$ such that they become different, whilst $\langle Q \rangle$ remains fixed.) these two planes are not parallel, and hence intersect each other on a straight line, which divided the plane Eq.~(\ref{dAve-Q^2}) into two halves.  As illustrated in Fig.~\ref{fig:aveq}, in one half, $d \langle Q^2 \rangle$ is positive, whereas in the other half, $d \langle Q^2 \rangle$ is negative.  All we need is to pick an arbitrary point $(dq_1, dq_2, dq_3)$ in the negative half plane, such that $d \langle Q^2 \rangle$  is negative.

Constructed as above, the redistribution process tunes the parameters $p_1, p_2, p_3, \sigma_3, q_1, q_2, q_3$ such that (i) Eq.~(\ref{FT-1}) and the normalization conditions (\ref{normalization-pdf}) remain valid, (ii) $p_1$ increases whereas $p_2$ decreases, (iii) the averages $\langle \Sigma \rangle$ and $\langle Q \rangle$ remain fixed, (iv) the variances $ \langle \delta \Sigma^2 \rangle$ and $ \langle \delta Q^2 \rangle$ decrease.  Doing this iteratively either $p_2$ or $p_3$ will eventually become zero, and we can remove the point  from the support of $p_N(\sigma,q)$ (and also remove the dual point from the support of $\bar p_N(\sigma, q)$).  As a consequence we obtain a pair of $(N-1)$-point distributions $p_{N-1}(\sigma,q), \bar p_{N-1}(\sigma,q)$, such that $p_{N-1}(\sigma,q)$ has the same averages $\langle \Sigma \rangle$ and $\langle Q \rangle$, but smaller variances $ \langle \delta \Sigma^2 \rangle$ and $ \langle \delta Q^2 \rangle$ comparing to $p_{N}(\sigma,q)$.  Keep doing this,  we eventually obtain a pair of 2-point distributions $p_{2}(\sigma,q), \bar p_{2}(\sigma,q)$, such that $p_{2}(\sigma,q)$ has the same averages $\langle \Sigma \rangle$ and $\langle Q \rangle$, but smaller variances $ \langle \delta \Sigma^2 \rangle$ and $ \langle \delta Q^2 \rangle$ comparing to original distribution $p_{N}(\sigma,q)$.

\appendix

\end{document}